# Photo-accelerated hot carrier transfer at MoS$_2$/WS$_2$: a first-principles study

*Zhi-Guo Tao*[1,2*], *Guo-Jun Zhu*[1,2*], *Weibin Chu*[1,2], *Xin-Gao Gong*[1,2], *and Ji-Hui Yang*[1,2†]

[1]Key Laboratory for Computational Physical Sciences (MOE), State Key Laboratory of Surface Physics, Department of Physics, Fudan University, Shanghai 200433, China

[2]Shanghai Qizhi Institution, Shanghai 200232, China

**ABSTRACT:** Charge transfer in type-II heterostructures plays important roles in determining device performance for photovoltaic and photocatalytic applications. However, current theoretical studies of charge transfer process don't consider the effects of operating conditions such as illuminations and yield systemically larger interlayer transfer time of hot electrons in MoS$_2$/WS$_2$ compared to experimental results. Here in this work, we propose a general picture that, illumination can induce interfacial dipoles in type-II heterostructures, which can accelerate hot carrier transfer by reducing the energy difference between the electronic states in separate materials and enhancing the nonadiabatic couplings. Using the first-principles calculations and the ab-initio nonadiabatic molecular dynamics, we demonstrate this picture using MoS$_2$/WS$_2$ as a prototype. The calculated characteristic time for the interlayer transfer (60 fs) and the overall relaxation (700 fs) processes of hot electrons is in good agreement with the experiments. We further find that illumination mainly affects the ultrafast interlayer transfer process but has little effects on the relatively slow intralayer relaxation process. Therefore, the overall relaxation process of hot electrons has a saturated time with increased illumination strengths. The illumination-accelerated charge transfer is expected to universally exist in type-II heterostructures.

---

[*] These authors contributed equally to this work.

[†] jhyang04@fudan.edu.cn

## I. INTRODUCTION

Heterostructures with type-II band alignments, in which the conduction band minimum (CBM) and valence band maximum (VBM) reside in two separate materials, are widely used for optoelectronic applications including solar cells and photodetectors[1, 2, 3, 4], because they can assist in the efficient separation of photoexcited electrons and holes. In type-II heterostructures, charge transfer at interfaces plays essential roles in determining device performance and therefore it is of great importance to understand charge transfer process. However, traditional heterostructures made of three-dimensional (3D) materials usually have lattice mismatch. As a result, the interfacial structures are very complicated with defects and large system sizes, making it difficult to study charge transfer process both experimentally and theoretically.

Thanks to the emerging of two-dimensional (2D) materials, Van der Waals (VDW) heterostructures made of different 2D materials with clean interfaces provide a simple and new platform for studying charge transfer process as well as exploring new devices such as ultrathin photodetectors[5, 6], tunneling transistors[7, 8, 9] and solar cells[10, 11]. Contrary to our conventional wisdom, although the layer coupling is weak in VDW type-II heterostructures, ultrafast interlayer charge transfers[12, 13, 14, 15] is observed. Wang et al.[12] experimentally reported that hot holes transfer from the $MoS_2$ layer to the $WS_2$ layer within 50 femtosecond (fs). Chen et al.[13] observed that hot electrons transfer from the $WS_2$ layer to the $MoS_2$ layer within 50 fs and then relax to the band edge after about 800 fs. Theoretically, Zhang et al.[16] used time-dependent density functional theory (TDDFT) to simulate the ultrafast transfer of holes and they reported a time scale of 100 fs. Zheng et al.[17] used ab-initio nonadiabatic molecular dynamics (NAMD) to simulate the relaxation of electrons and they found that about 3/4 of the electrons relax to the band edge after 1000 fs. We find that theoretically simulated charge transfer time is systematically longer than the experimental results. Although there could be many reasons for the discrepancy between experiments and theories such as substrates, interfacial defects, inaccurate DFT band structures of ground states, etc., one fact that photo-induced electrons and holes will definitely affect the electronic band structures at the interface, is not considered in previous theoretical works. Consequently, whether and how illumination plays roles in the process of interfacial charge transfer is not known, revealing which is not only important for understanding light-matter interaction but also will benefit optoelectronic performance of many 2D and 3D devices.

In this Letter, for the first time, we investigate the effect of illumination on the ultrafast transfer and relaxation of hot electrons at type-II heterostructures. We propose a general picture that, the photo-induced electrons and holes residing in different materials of type-II heterostructures will form an interfacial dipole, which reduces band energy difference in separate materials, thus accelerating hot electron transfer. We demonstrate our picture in $WS_2/MoS_2$ heterostructure. Using the first-principles calculation methods and NAMD simulation techniques, we systematically explore the dynamics of hot electrons under different illumination strengths. Our simulation results indicate that the transfer of hot electrons is indeed accelerated under illumination and the calculated transfer time is in excellent agreement with experimental results[13]. We expect that illumination-accelerated charge transfer is universal in both 2D and 3D type-II heterostructures.

## II. COMPUTATIONAL DETAILS

The density functional theory (DFT) calculations are performed using the Vienna ab initio simulation package (VASP)[18, 19, 20]. The electron and core interactions are treated using the frozen-core projected augmented wave (PAW) approach[21]. For all the calculations, the generalized gradient approximation (GGA) formulated by Perdew, Burke, and Ernzerhof (PBE)[22] is adopted. The van der Waals (vdW) interactions are considered using the DFT-D2[23] method. To simulate the steady illumination conditions, the constrained occupation method[24, 25] is used, in which fractional electrons are removed from the VBM to the CBM of the whole heterostructure. For the calculations of electronic structures, an energy cutoff of 400 eV and a 18x18x1 k-point mesh are adopted. To study the charge transfer process, we perform the non-adiabatic molecular dynamic simulations (NAMD) as implemented in the Hefei-NAMD code[26]. Γ point is used in the ab-initio molecular dynamic (AIMD) simulations for an orthogonal $3 \times 3 \times 1$ supercell of $MoS_2/WS_2$ with 108 atoms. We sample $1\times10^4$ trajectories for 1ps with a time step of 1fs. The NAMD results are based on averaging over 100 different initial configurations. The nonadiabatic couplings are calculated using CAnac with phase correction and state tracking[27, 28, 29, 30].

### III. RESULT AND DISCUSSION

The band diagram of a typical type-II heterostructure is shown in Fig. 1. Before illumination, the CBM and VBM of material B are higher than the CBM and VBM of material A, respectively. The band offsets are $\Delta E_c$ and $\Delta E_v$ for the CBM and VBM, respectively. Under illumination, electrons are excited from the valance band to the conduction band in either A or B or both, depending on the photon energies and band gaps. Driven by thermodynamic effects, the excited carriers will finally relax to the band edges of the whole heterostructure, i.e., electrons will relax to the CBM of A and holes will relax to the VBM of B. Consequently, an interfacial dipole will form, which will push the bands in A upwards and pull the bands in B downwards, respectively, thus reducing the energy differences between bands in A and bands in B, i.e., $\Delta E_c$ and $\Delta E_v$ will be reduced. In this case, if there is a hot electron in B, it will be easier for it to transfer to A and interlayer charge transfer will be accelerated. Such a picture is expected to commonly exist in type-II heterostructures.

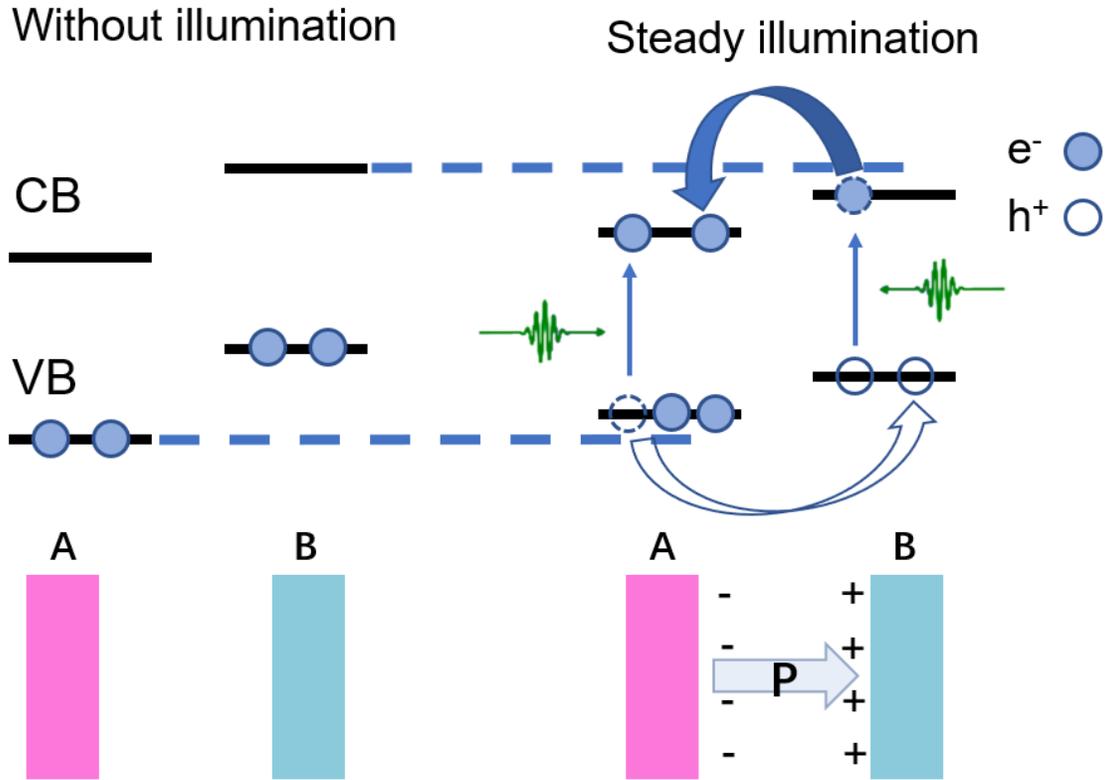

FIG. 1. A schematic diagram to show the photoexcitation and carrier transfer in an A/B heterostructure with/without illumination. The dashed circle represents carriers transferring from A(B) to B(A)

To demonstrate the above picture and quantitatively determine the effects of photo-induced interfacial dipoles on charge transfer at type-II heterostructures, we theoretically study the charge transfer at $MoS_2/WS_2$ with considering the illumination conditions using NAMD. Before presenting the simulation results, we first discuss the band structure of $MoS_2/WS_2$ heterostructure. Here we consider the C7 stacking of $MoS_2$ and $WS_2$ as it is previously reported to be the most stable[31, 32]. In MoS$_2$ or WS$_2$ monolayer, both the VBM and CBM states lie at the K point. When they are stacked, the band edges at the K point are distributed in separate materials. As seen in Fig. 2, the lowest conduction band edge at the K point mainly resides in MoS$_2$ (denoted as CB@K_MoS$_2$) while the higher conduction band edge at the K point mainly resides in WS$_2$ (denoted as CB@K_WS$_2$). Similarly, the higher (lower) valance band edge at the K point mainly resides in WS$_2$ (MoS$_2$). Such kind of band edge distributions clearly indicates that the system is a type-II heterostructure. We note that, the overall VBM of the heterostructure lies at the Γ point, compared to that the VBM in MoS$_2$ or WS$_2$ monolayer is located at the K point. The difference mainly results from the interlayer orbital hybridization due to interlayer interactions, as we find that the VBM is distributed in both MoS$_2$ and WS$_2$ (see Fig. 2). Because this state resides more in WS$_2$, we denote it as VB@Γ_WS$_2$. Our calculated band structure agrees well with previous studies[17, 33].

Upon illumination by low-energy photons which have energies slightly larger than the bandgaps of MoS$_2$ or WS$_2$, electrons will be excited from the valence band edges to the conduction band edges in MoS$_2$ and/or in WS$_2$. Then the hot electrons (holes) will relax and transfer to the CB@K_MoS$_2$ (VB@Γ_WS$_2$) finally. Here we mainly focus on the transfer and relaxation of hot electrons from the CB@K_WS$_2$ to the CB@K_MoS$_2$. Note that, between the CB@K_WS$_2$ and CB@K_MoS$_2$, there is an intermediate state which is located at the T valley. Electronic character analysis shows that this state is mainly contributed by the electronic state of MoS$_2$ with a small part distributed in WS$_2$, indicating that it's a hybridized state. Nevertheless, we still denote this state as CB@T_MoS$_2$. Previous studies show that such as intermediate state can help the electron transfer and relaxation following the route of CB@K_WS$_2$−>CB@T_MoS$_2$−>CB@K_MoS$_2$ while it is difficult for hot electrons to directly transfer from CB@K_MoS$_2$ to CB@K_WS$_2$. Their results show that it takes more than 1000 fs for hot electrons to relax from CB@K_WS$_2$ to CB@K_MS$_2$, which is longer than the experimental results (800 fs), as no illumination conditions are considered.

Once illumination is applied, hot carrier transfer and relaxation happens. Since the overall CBM and VBM of the heterostructure are distributed in different layers, holes will be enriched in WS$_2$ while electrons will be enriched in MoS$_2$. Therefore, we can simulate the illumination conditions by constraining the electron occupations at different states. At the room temperature, the photo-excited carrier densities[12, 13] in the MoS$_2$/WS$_2$ heterostructure is about $10^{12}$–$10^{13}$ cm$^{-2}$, which corresponds to exciting about 0.1 electrons from the VB@Γ_WS$_2$ to the CB@K_MoS$_2$ in our simulated 3x3x1 supercell. To study the illumination effects on hot electron transfer and relaxation, here we consider three illumination conditions (α = 0, 0.05, 0.1), which correspond to the photo-excited carrier densities of 0, $3.16 \times 10^{12}$, and $6.32 \times 10^{12}$ cm$^{-2}$, respectively.

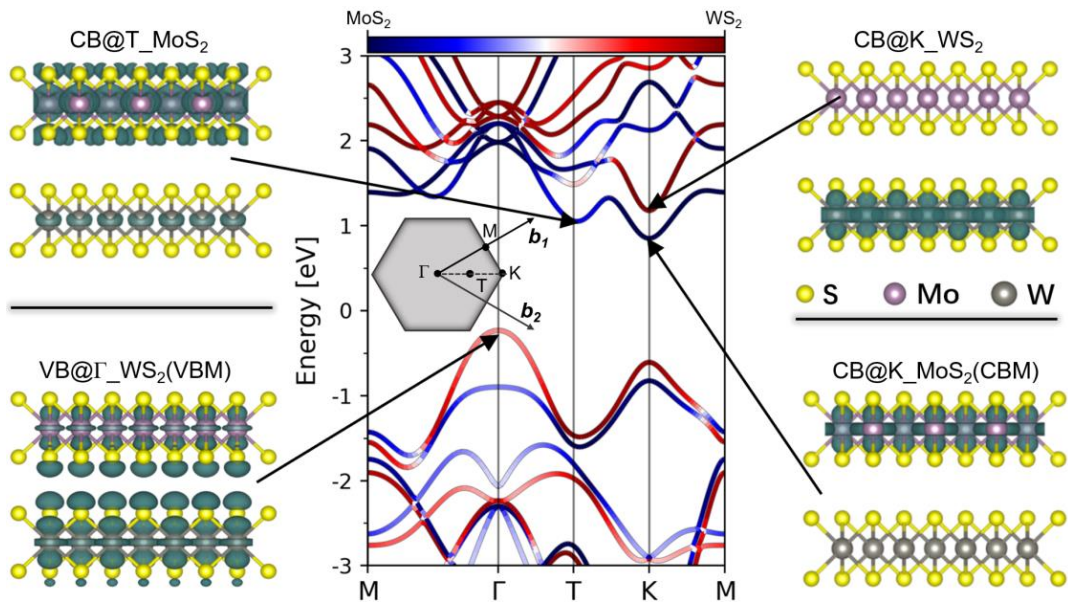

FIG. 2. Band structure of the MoS$_2$/WS$_2$ heterostructure and the electronic characters of considered states. The Brillion zone is shown in the inset. The color map indicates the orbital contributions from different materials, i.e., red for WS$_2$ and blue for MoS$_2$.

Now we consider the hot electron transfer and relaxation under different illumination conditions. As mentioned above, the total relaxation path of hot electrons, that is, CB@K_WS$_2$−>CB@T_MoS$_2$−>CB@K_MoS$_2$, can be divided into two processes: hot electron transfer between layers (CB@K_WS$_2$−>CB@T_MoS$_2$) and electron relaxation in MoS$_2$ (CB@T_MoS$_2$−>CB@K_MoS$_2$) after transfer. Using the NAMD simulations, we can obtain the transfer and total relaxation time of hot electrons. Our results are shown in Fig. 3. Initially, one hot electron is put at the CB@K_WS$_2$ state. If more than 0.5 electrons evolve to another state with time, then we say the transfer or the relaxation occurs. Without illumination, our results show that it takes near 270 fs for the hot electron to transfer from the CB@K_WS$_2$ to the CB@T_MoS$_2$ and the total relaxation time from the CB@K_WS$_2$ to the CB@K_MoS$_2$ state is about 1000 fs (see Fig. 3(a)). While our results are similar to previous theoretical studies[17], the time values are larger than the experimental measurements[13]. When the illumination strength corresponds to the photoexcitation of 0.05 electrons between the overall band edges of the heterostructure, the transfer and overall relaxation time of the hot electron are 100 fs and 700 fs, respectively (see Fig. 3(b)). As the illumination strength further increases corresponding to the photoexcitation of 0.1 electrons, the transfer time of hot electrons further reduces to about 60 fs while the overall relaxation time is barely reduced further. We note that, these two characteristic time values under illumination conditions are in good agreement with the experiment[13]. The discrepancies between previous theoretical results and experimental values are thus due to the illumination conditions in our opinions.

We further simulate the interlayer charge transfer process CB@K_WS$_2$−>CB@T_MoS$_2$ and the intralayer charge relaxation process CB@T_MoS$_2$−>CB@K_MoS$_2$ separately to obtain more understandings about the illumination effects on hot electron behaviors at the heterostructure. As shown in Fig. S1, our results show that illumination can accelerate the interlayer transfer of hot electrons but has negligible effects on the intralayer relaxation with the time scale of the second process much longer than that of the first one. Consequently, the main bottleneck of the overall hot electron relaxation process is actually the intralayer relaxation. This explains why the overall relaxation of the hot electron tends to saturate with increased illumination strengths.

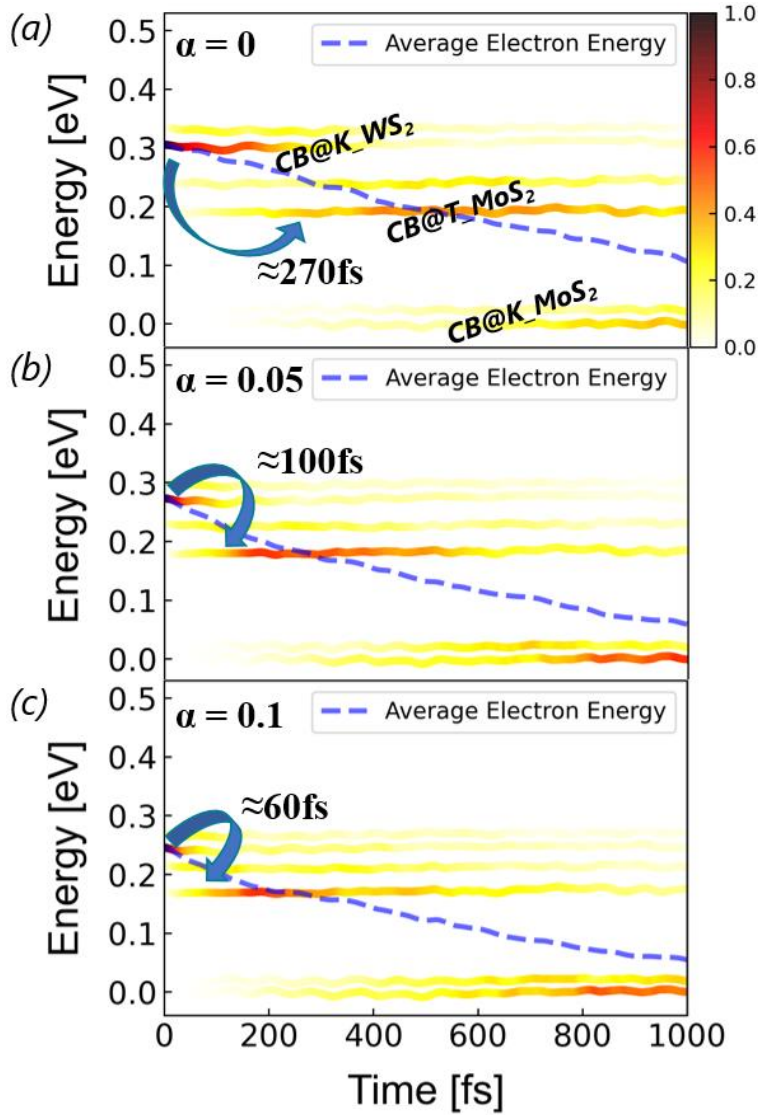

FIG. 3. Simulation results for the evolution of a hot electron from the CB@K_WS$_2$ state to the CB@K_MoS$_2$ state. (a), (b), and (c) Time evolutions of the electron occupations and averaged electron energy at 300 K under different illumination strengths. The colorful strips indicate the electron occupations of relevant energy states as time evolves, and the dashed line represents the averaged electron energy. The energy reference is the averaged CBM energy.

To understand why illumination accelerates the interlayer transfer of the hot electron, we turn to analyze the carrier dynamics between different energy states determined by the nonadiabatic coupling (NAC) term[34, 35, 36, 37, 38, 39, 40], which is given by

$$d_{ij} = \left\langle \varphi_i \left| \frac{\partial}{\partial t} \right| \varphi_j \right\rangle = \frac{\langle \varphi_i | \nabla_R H | \varphi_j \rangle}{\epsilon_j - \epsilon_i} \dot{R},$$

where H is the Kohn–Sham Hamiltonian, $\varphi_i, \varphi_j, \epsilon_i\ and\ \epsilon_j$ are the wave functions and eigenvalues for the electronic states j and k, respectively, and $\dot{R}$ is the velocities of nuclei. Note that, the probability of electron transition is proportional to the square of NAC and larger NAC indicates faster electron transitions between two electronic states. Our results in Fig. 4(a) show that, the averaged NAC between the CB@K_WS$_2$ and CB@K_MoS$_2$ states is small compared to the one between the CB@K_WS$_2$ and CB@T_MoS$_2$ states as well as the one between the CB@T_MoS$_2$ and CB@K_MoS$_2$ states, suggesting that the direct electron transitions from the CB@K_WS$_2$ to the CB@K_MoS$_2$ is very difficult while the combined interlayer transfer and intralayer relaxation, i.e., CB@K_WS$_2$−>CB@T_MoS$_2$−>CB@K_MoS$_2$, is the dominate mechanism for the electron relaxation at the heterostructure. Our analysis is consistent with previous works[17].

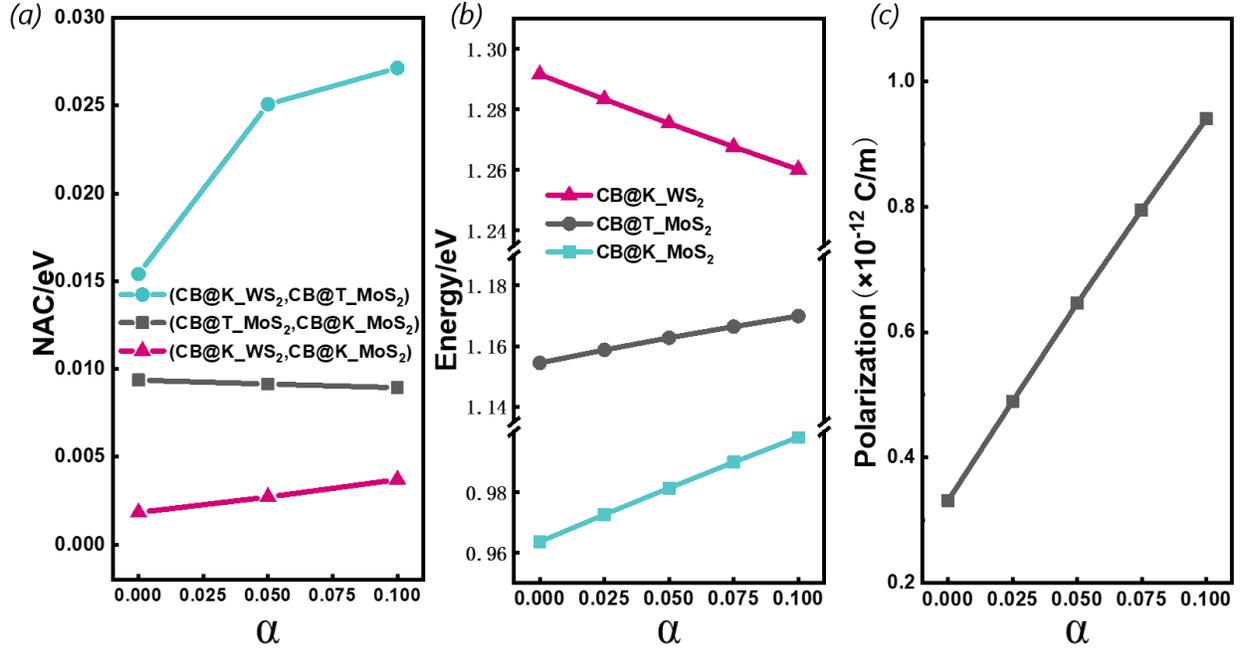

FIG. 4. Interfacial carrier-dynamics-related properties dependence on illumination strengths. (a) The averaged absolute values of the NAC between considered electronic states under different illumination strengths. (b) and (c) The energy levels and interfacial dipole dependence on illumination strengths, respectively.

Now we see how illumination affects the NAC terms. As shown in Fig. 4(a), the averaged NAC between the CB@K_$WS_2$ and CB@T_$MoS_2$ states increase with the increase of light intensity. In contrast, the NAC between the CB@T_$MoS_2$ and CB@K_$MoS_2$ states is little affected by the illumination. The changes of the NAC terms under the illumination conditions thus explain the accelerated interlayer charge transfer as well as the nearly unaffected intralayer charge relaxation.

To understand the underlying mechanism behind the illumination effects on the NAC terms, we note that, the NAC value is inversely proportional to the energy difference between the two states and is proportional to the atomic velocities as well as the electron−phonon couplings. The electron−phonon couplings can be reflected by the energy fluctuations of the Kohn−Sham states in the molecular dynamical simulation (see Figs. S2(a)-S2(c)) or more clearly by the corresponding

Fourier transform (FT) spectra[35, 36] (see Figs. S2(d)-S2(f)). As we can see, the amplitudes of the three conduction band energy fluctuations are basically the same under different illumination conditions, indicating that the electron−phonon coupling strength is little affected by illuminations. This is understandable because in our simulation process, the temperature of the system is maintained as 300 K and therefore the illumination effect on atomic vibrations should be negligible if no heating effect is considered. Similarly, the velocities of nuclei are also expected to have little changes. Consequently, illumination affects the NAC terms mainly through affecting the band energy differences. Indeed, from Figs. S2(a)-S2(c), we can see that, there are clear changes of the energy differences between $MoS_2$ and $WS_2$ bands with varying illumination strengths.

To study how the energy level change with illumination strengths, we calculate the electronic structures of the heterostructure under the illumination conditions using the equilibrium structure. As shown in Fig. 4(b), with the illumination strengths increasing, the energy levels of the CB@T_$MoS_2$ and CB@K_$MoS_2$ states are increased, while the energy level of the CB@K_$WS_2$ state is decreased. According to Fig. 1, we know that under the illumination conditions, electrons are enriched in $MoS_2$ and holes are enriched in $WS_2$, inducing the dipole across the heterostructure. The stronger the illumination strength, the larger the dipole (as shown in Fig. 4(c)), which reduces the electrostatic potential in $MoS_2$ and increases that in $WS_2$, leading to the increase of the energy level of the $MoS_2$ layer and the decrease of the energy level of the $WS_2$ layer. Accordingly, the energy level difference between the CB@K_$WS_2$ and CB@T_$MoS_2$ changes from 137 meV when there is no illumination to 90 meV when the illumination strength is 0.1 (see Fig. 4(b)), which is reduced by as much as 1/3. Consequently, the NAC term increases, leading to the accelerated interlayer electron transfer. In contrast, the energy difference between the CB@T_$MoS_2$ and CB@K_$MoS_2$ is only slightly reduced because the CB@T_$MoS_2$ is a hybrid state with a small part of it distributed in WS$_2$. In addition, with the increased illumination, the distribution of CB@T_MoS$_2$ orbit in WS$_2$ layer is slightly increased (see Table S1), which might weaken the coupling between CB@T_MoS$_2$ and CB@K_MoS$_2$. The consequence is that the NAC term between the CB@T_MoS$_2$ and CB@K_MoS$_2$ is nearly unchanged under illumination and therefore the intralayer electron relaxation is little affected. Note that, although the energy level difference between the CB@K_WS$_2$ and CB@K_MoS$_2$ has a large decrease under illumination, the relatively large energy difference always leads to the very small NAC term whether the system is under illumination or not. Consequently, the direct electron transition from the CB@K_WS$_2$ to the CB@K_MoS$_2$ can be reasonably ignored.

Here we have used the simple MoS$_2$/WS$_2$ system as an example to demonstrate the general effects of the illumination on the charge transfer at a type-II heterostructures. Note that, the ultrafast transfer of hot electrons from the WS$_2$ layer to the MoS$_2$ layer via the CB@T_MoS$_2$ energy level indicates that, a hybridized intermediate state which is located in between the band edge states of two materials in a type-II heterostructure, can accelerate carrier transfer and enhance carrier separations, especially under illumination conditions. Consequently, the performance of optoelectronic devices might be improved through creating such intermediate states, i.e., by inserting some window layer in type-II heterostructures. Another thing we want to point out is that, although the interfacial dipole enhances the electron transfer by reducing band energy differences in separated layers, electrons in MoS$_2$ must be extrapolated in time. Otherwise, accumulated electrons in MoS$_2$ will increase the dipole field, thus preventing interfacial electron transfer. Under very strong illuminations, i.e., in concentrated solar cells, once the extrapolation speed is surpassed by the transfer speed, the device efficiencies might be reduced[41, 42]. Therefore, by manipulating illumination strengths to make the charge transfer and carrier extrapolation processes coordinate, device performance could be enhanced.

## IV. CONCLUSION:

In summary, for the first time, we have investigated the effect of illumination on the ultrafast transfer and relaxation of hot electrons in $MoS_2/WS_2$ heterostructure. We have demonstrated the picture that, the interfacial dipole induced by illumination can reduce the energy difference between the states in $MoS_2$ and those in $WS_2$, thus accelerating interlayer transfer of hot electrons by enhancing the nonadiabatic couplings. We expect this picture can be widely applied to understand the carrier transfer in both 2D and 3D type-II heterostructures. In additions, based on our studies, we propose that it might offer a feasible strategy to improve the performance of optoelectronic devices by manipulating illumination strengths to have coordinate charge transfer and carrier extrapolation processes.

## ACKNOWLEDGMENT

This work was supported in part by National Natural Science Foundation of China Grant No. 12188101, the Special Funds for Major State Basic Research (2022YFA1404603), National Natural Science Foundation of China Grant No. 11991061 and 11974078). Computations were performed at the High-Performance Computing Center of Fudan University.

# Supporting Information

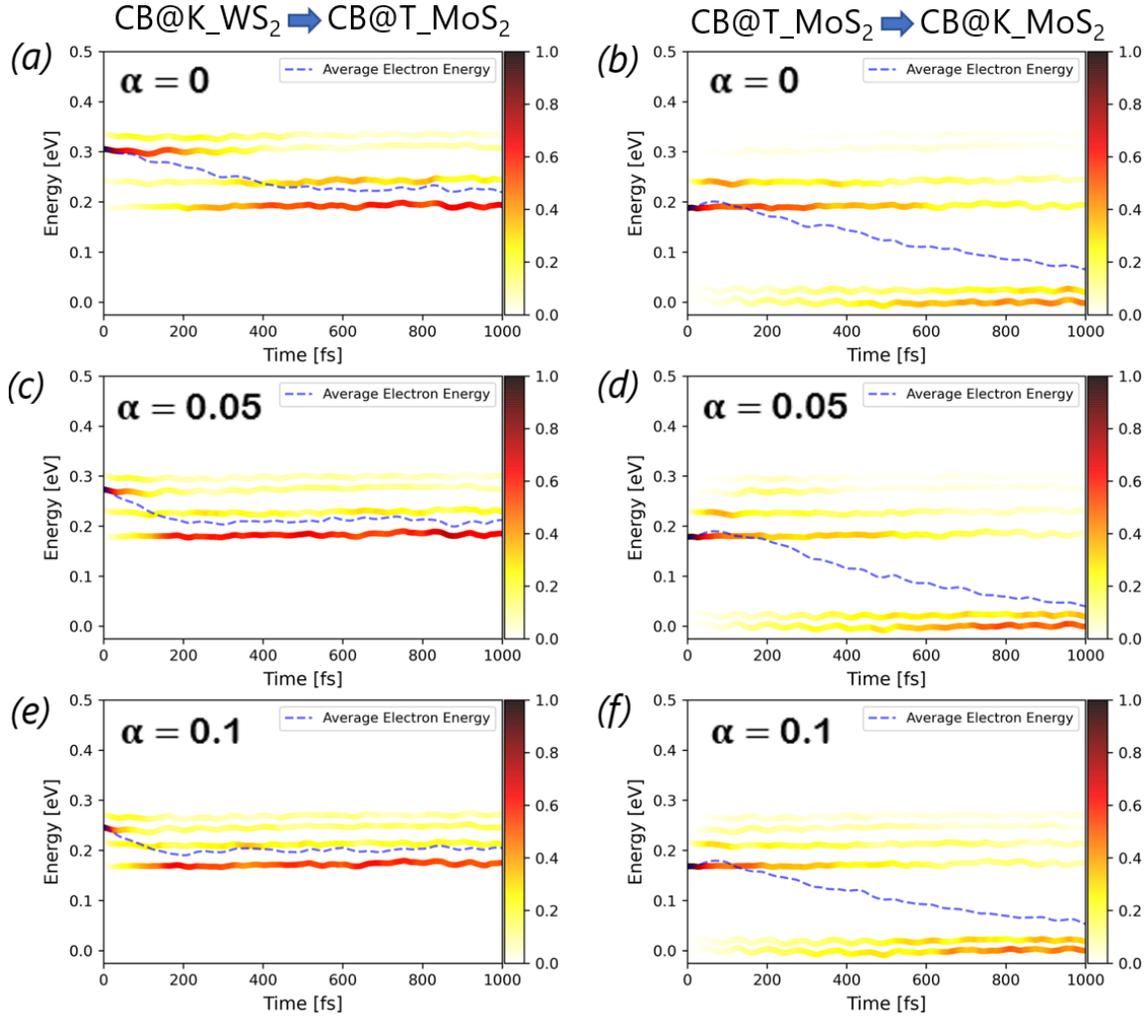

FIG. S1. Time-dependent electron energy change at 300K under different light intensities. (a), (c), (e) corresponds to the transfer of hot electrons from CB@K_$WS_2$ to CB@T_$MoS_2$. (b), (d), (f) corresponds to the transfer of hot electrons from CB@T_$MoS_2$ to CB@K_$MoS_2$. The color strips indicate the electron distribution on different energy states, and the dashed line represents the averaged electron energy. The energy reference is the average CBM energy.

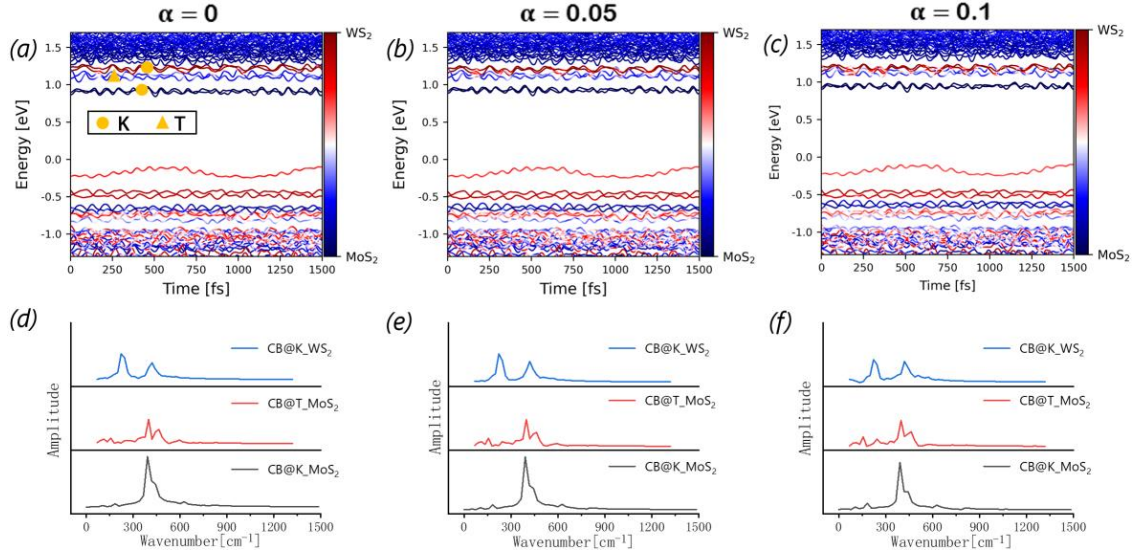

FIG. S2. Time evolutions of the energy states near CBM(a-c) and their FT spectrum(d-f). The triangle and circle in panels a indicate the momentum of different energy states.

TABLE S1. Orbital spatial distributions of CB@T_MoS$_2$ under different illumination

|  | α=0 | α=0.05 | α=0.1 |
|---|---|---|---|
| WS2 | 22.2% | 23.7% | 26.3% |
| MoS2 | 77.8% | 76.3% | 73.6% |